# A polarizable ion model for the structure of molten AgI

Vicente Bitrián and Joaquim Trullàs[a)]

Departament de Física i Enginyeria Nuclear, Universitat Politècnica de Catalunya, Campus Nord UPC B4-B5, 08034 Barcelona, Spain.

Moises Silbert

School of Mathematics, University of East Anglia, Norwich NR4 7QF, United Kingdom.

[a)] Author to whom correspondence should be addressed. E-mail: quim.trullas@upc.edu

The results are reported of the molecular dynamics simulations of the coherent static structure factor of molten AgI at 923 K using a polarizable ion model. This model is based on a rigid ion potential, to which the many body interactions due to the anions induced polarization are added. The calculated structure factor is in better agreement with recent neutron diffraction data than that obtained by using simple rigid ion pair potentials. The Voronoi-Delaunay method has been applied to study the relationship between voids in the spatial distribution of cations and the prepeak of the structure factor.

AgI has long been regarded as the paradigmatic superionic conductor. It experiences a sudden increase in its ionic conductivity of around three orders of magnitude, at 420 K, known as the β-α transition.[1] Howe *et al.*,[2] following their diffuse neutron scattering from α-AgI, noted the similarity between this structure to that of molten CuCl.[3,4] This result prompted Stafford and Silbert[5] to carry out theoretical calculations of the pair distribution functions and partial structure factors of molten AgI within the hypernetted chain approximation. They used the same effective pair potentials that Parrinello *et al.*[6] had successfully used to account for the β-α transition, and suggested that their approximation may reproduce correctly the general structural features of the melt. Shortly after the calculations by Stafford and Silbert, the first experimental data of the structure factor of molten AgI were published,[7,8] and it was found that the theoretical calculations of the static structure factor $S(k)$ were in fair agreement with experiment.[7] Moreover, MD simulations of the ionic transport properties suggested that molten AgI, near melting, retains the superionic character of the α-phase.[9,10] However, the calculated $S(k)$ failed to reproduce the almost featureless broad main peak between the shoulder at 1.8 Å$^{-1}$ and the maximum at around 2.8 Å$^{-1}$, as well as the prepeak, or first sharp diffraction peak, which appears at around 1 Å$^{-1}$. Actually no prepeak was expected in the calculations as these were carried out using rigid ion pairwise potentials.

Some time ago Madden and co-workers suggested that the induced polarization contribution is essential to understand some of the features and properties of ionic systems.[11] This was particularly evident in their work on the divalent and trivalent metals chloride melts.[11,12] One of their earlier works was their study of AgCl, when they suggested that both the dipole and quadrupole contributions were required.[13] However, we showed that, in the molten phase, all the features of its structure were reproduced if the anion induced polarization contributions are added to the Vashishta-Rahman potentials.[14] This approach had the advantage that the values for the diffusion coefficients were higher than those in Ref. 13, and the value for the ionic conductivity was in better agreement with experiment. More recently a thorough study of polarizable ion models in molten AgBr has also been carried out.[15]

In the present work we have used rigid ion pair potentials of the form proposed in Ref. 6, which we write

$$f_{ab}(r) = f^0_{ab}(r) - \frac{P_{ab}}{r^4} \qquad (1)$$

with

$$f_{ab}^0(r) = \frac{z_a z_b e^2}{r} + \frac{H_{ab}}{r^{h_{ab}}} - \frac{C_{ab}}{r^6} \quad (2)$$

The first term on the rhs of equation (2) is the Coulomb interaction between the charges, with $z_a < 1$ the effective charge in units of the fundamental charge $e$; the second models the repulsion between the ions, with $H_{ab} = A(s_a + s_b)^{h_{ab}}$, where the $s_a$ are related to the ionic radii, $A$ defines the strength and $h_{ab}$ the hardness of the repulsive interactions. The third term is the van der Waals contribution, with $C_{ab} = (3/2)a_a a_b (E_a^{-1} + E_b^{-1})^{-1}$, where the $a_a$ are the polarizabilities and the $E_a$ are related to the ionization potentials of the cations and electron affinities of the anions. The second term on the rhs of equation (1) denotes the effective monopole-induced dipole interaction with $P_{ab} = (1/2)(a_a z_b^2 + a_b z_a^2)e^2$.

The polarizable ion model used in this work is constructed by adding the induced polarization contributions to the pair potential $f_{ab}^0(r)$. We assume that, on an ion placed at position $\mathbf{r}_i$, the local electric field $\mathbf{E}_i$ due to all the other ions, induces a dipole whose moment $\mathbf{p}_i = a_i \mathbf{E}_i$. The local field may be written as $\mathbf{E}_i = \mathbf{E}_i^q + \mathbf{E}_i^p$, where $\mathbf{E}_i^q$ is the field at $\mathbf{r}_i$ due to all the point charges except the charge $q_i = z_i e$ at $\mathbf{r}_i$, and $\mathbf{E}_i^p$ the field at $\mathbf{r}_i$ due to all the dipole moments except $\mathbf{p}_i$. The potential energy of this polarizable model may be written as

$$U = \frac{1}{2}\sum_{i=1}^N \sum_{j \neq i}^N f_{ij}^0(r_{ij}) - \sum_{i=1}^N \mathbf{p}_i \cdot \mathbf{E}_i^q - \frac{1}{2}\sum_{i=1}^N \mathbf{p}_i \cdot \mathbf{E}_i^p + \sum_{i=1}^N \frac{p_i^2}{2a_i} \quad (3)$$

It is possible to construct other polarizable ion models with a short-range damping polarizability that opposes the electrically induced dipole moments; these are discussed in Ref. 15.

For the parameterization of $f_{ab}(r)$ we have chosen the values given by Shimojo and Kobayashi.[16] They showed that, using these parameters the α-phase properties are reproduced at the appropriate thermodynamic state. Following Ref. 16 we have assumed an effective charge $|z| = 0.5815$ and the polarizabilities $a_- = 6.12$ Å$^3$ and $a_+ = 0$. Then, $P_{++} = C_{++} = C_{+-} = 0$. The other parameters are $h_{++} = 11$, $h_{+-} = 9$, and $h_{--} = 7$; $H_{++} = 0.162$ eVÅ$^{11}$, $H_{+-} = 1309.1$ eVÅ$^9$, and $H_{--} = 5325.2$ eVÅ$^7$; $P_{--} = 2P_{+-} = 29.796$ eVÅ$^4$ and $C_{--} = 84.4$ eVÅ$^6$.

In the present work we have carried out MD simulations over $3 \times 10^5$ time steps, with a time step $\Delta t = 5 \times 10^{-15}$ s, using $N = 1000$ ions placed in a cubic box of side $L$ with an ionic density $r = N/L^3 = 0.0281$ Å$^{-3}$. The temperature is 923 K, the same at which experimental data were measured, about 100 K higher than that of the melting point (825 K). According to Janz et al.[17] the ionic density at this temperature is that used in this work. Computational details, as well as the prescription to calculate the structure factors, are described in Refs. 14 and 15. From the partial structure factors, $S_{ab}(k)$, we have calculated the coherent structure factor, $S(k) = [b_+^2 S_{++}(k) + b_-^2 S_{--}(k) + 2b_+b_- S_{+-}(k)]/(b_+^2 + b_-^2)$,[18] using the neutron scattering lengths values $b_+ = b_{Ag} = 5.922$ fm and $b_- = b_I = 5.280$ fm.[19]

The experimental neutron diffraction data, with which we compare our simulation results, are the results of the Kyushu University group.[20] These data agree with those published by Shirakawa et al.,[21] who reanalyzed the original diffraction data of Inui et al.[8] The $S(k)$ obtained by the Kyushu University group and Shirakawa et al. exhibit the same featureless broad main peak with a shoulder at 1.8 Å$^{-1}$ and the maximum at 2.8 Å$^{-1}$ (see Fig.1). Moreover, it has a prepeak at around 1 Å$^{-1}$, which is more pronounced in the Kyushu University data.

The $S(k)$ results of our simulations using the polarizable ion model (PIM) are compared with experimental data in Fig. 1. For the sake of completeness, we also include the results obtained using the rigid ion model (RIM), equation (1). We note that the RIM results for $S(k)$ do not show the prepeak and, while the wave number of the first peak agrees with that of the experimental shoulder at the beginning of the main broad peak, the maximum is slightly displaced towards a larger $k$ than that of the experimental maximum at 2.8 Å$^{-1}$. As a result the broad, almost featureless, contribution in the middle becomes a trough. Turning to the PIM results, we note that they reproduce the prepeak and improve the structure of the broad main peak with the highest value in agreement with experiment. Nevertheless, it appears that the simulations tend to magnify the initial shoulder, as well as the mid features with a shoulder in place of the RIM trough. Moreover, beyond the main peak the $S(k)$ oscillations damp faster than experiment, suggesting that a better agreement with experiment may result if more elaborate pair potentials are used in the simulations in place of the relatively simple potentials of the Vashishta-Rahman form.

The partial structure factors, $S_{ab}(k)$, and the radial distribution functions, $g_{ab}(r)$, calculated from simulations are shown in Fig. 2 and Fig. 3. It is interesting to note that, although we only take into account the anion polarizability, it is the PIM cation-cation structure that is mainly affected by the induced polarization interactions, while the anion-anion structure is practically unaffected. The most salient effect of the anion polarizability is the prepeak in $S_{AgAg}$. Furthermore, the $S_{++}$ for the PIM shows a second peak at about 2.25 Å$^{-1}$ and its first peak is lower than that for the RIM. Regarding the radial distribution functions, the first peak of the $g_{AgAg}$ for the PIM is shifted inwards with a deeper cations penetration, i,e, the separation between neighboring cations can be smaller than it would be the case if the anions were not polarized. This features may be attributed to the screening of the cations repulsion due to the anion induced dipoles, namely, the negative ends of the anion dipoles attract the cations.

In the light of the above results, it seems that anion polarizability is in the origin of the prepeak in the structure factor. This agrees with the fact that the intensity of the prepeak in experimental structure factors of molten cuprous and silver halides (CuX and AgX, X=Cl, Br, I) increases with anion polarizabilty.[8,22] Salmon, in his analysis of the experimental data of molten and glassy 2:1 systems,[23] states that the prepeak (or "first sharp diffraction peak") appears to be a signature of directional bonding, present when the system comprises polarizable species or covalent bonds. The same conclusion may be drawn from the computer simulation studies on $ZnCl_2$ made by Madden and Wilson.[24,25] They proved that only if polarization effects are taken into account a prepeak which agrees well with that observed experimentally is obtained.

The prepeak is the signature of a new characteristic length scale in the pair correlations, larger than the distance between neighboring ions. Since it is the result of subtle intermediate-range correlations, its structural origin is difficult to distinguish in real-space, namely, the differences between the Fourier transform of $S(k)$ with and without the prepeak are barely perceptible. To gain more insight into the intermediate-range order related to the prepeak, Madden and Wilson proposed to consider the voids between atoms, but not the atoms themselves.[25] In order to define the voids in a (disordered) ionic configuration, they used the Voronoi-Delaunay method,[26,27] which defines the set of Delauney tetrahedra (DT). This set is such that every DT has one ion at every vertex, with no other atom centre lying within its circumsphere. Therefore, the circumsphere radius gives a measure of the empty space between the four atoms at the vertices of the DT. Each tetrahedron's face is shared with another DT, and the set of all tetrahedra fills the space without any gap.

Since the prepeak in $S(k)$ is due to the contribution of $S_{AgAg}(k)$, we have applied the Voronoi-Delaunay method to the cation configurations. Furthermore, since the anions are more closely packed than the cations, the formation of voids between the former is not expected. In Fig. 3/4, the circumsphere radii (or void radii, $R_V$) distribution, $D(R_V)$, is shown for the PIM and RIM. The former distribution is broader than the later, and peaks at a larger radius value. That is, the average cation configuration in the PIM has bigger empty spaces (voids), and comprises a wider variety of small and big voids. The induced dipoles tend to screen the repulsion between neighboring cations, which can approach at smaller distances, and low cation density regions (voids) are opened up. In order to find out whether there is a characteristic length scale in the spatial distribution of voids, we have also calculated the void-void structure factor,

$$S_{VV}(k) = \left\langle \frac{1}{N_V} \sum_{i=1}^{N_V} \sum_{j=i}^{N_V} \exp(i\mathbf{k}\cdot\mathbf{R}_{ij}) \right\rangle = \left\langle \frac{1}{N_V} \sum_{i=1}^{N_V} \exp(i\mathbf{k}\cdot\mathbf{R}_i) \sum_{j=1}^{N_V} \exp(-i\mathbf{k}\cdot\mathbf{R}_j) \right\rangle \quad (4)$$

where $N_V$ is the number of voids in the configuration, and $\mathbf{R}_{ij} = \mathbf{R}_i - \mathbf{R}_j$ is the vector joining the centres of the voids $i$ and $j$ centred at $\mathbf{R}_i$ and $\mathbf{R}_j$.

As can be seen in Fig. 2, the $S_{VV}(k)$ for RIM has the main peak at about 1.5 Å$^{-1}$, close to the wave number of the first peak of the RIM $S_{AgAg}(k)$, while the main peak of the $S_{VV}(k)$ for PIM is at $k_{VV} = 1.1$ Å$^{-1}$, very close to the wave number of the prepeak of the PIM $S_{AgAg}(k)$. Since the length scale $2\pi/k_{VV}$ is related to the distance between voids, the prepeaks in $S(k)$ and $S_{AgAg}(k)$ for PIM appears to be the signature of a periodicity of low cation density zones (or voids), due to anion polarization effects.

Summing up, we have shown in this paper that the main trends of the experimental structure factor of molten AgI at 923 K may be reproduced by adding the anion polarizability to simple rigid ion pair potentials. Furthermore, the structural origin of the prepeak has been related to a new length scale which characterizes the ordering of the voids between cations.

This work was supported by the DGICYT of Spain (Grant No. BFM2003-08211-C03-01), the DURSI of the Generalitat of Catalonia (Grant No. 2005SGR-00779) and the European Union FEDER funds (Grant No. UNPC-E015). One of us (V.B.) thanks the Ministry of Education of Spain for the FPU Grant No. AP2003-3408.

**Figure captions**

FIG. 1. Coherent static structure factors, $S(k)$, for molten AgI at 923 K from experimental data (open circles),[20] and MD results using the RIM (dotted line) and the PIM (solid line).

FIG. 2. MD results of the partial structure factors for the RIM (top) and the PIM (middle): $S_{AgAg}(k)$ (solid line); $S_{II}(k)$ (dashed line); and $S_{AgI}(k)$ (dotted line). Bottom: MD results of the void-void structure factor, $S_{VV}(k)$, for the RIM (dotted line) and the PIM (solid line).

FIG. 3. MD results of the radial distribution functions for the RIM (top) and the PIM (bottom): $g_{AgAg}(r)$ (solid line); $g_{II}(r)$ (dashed line); and $g_{AgI}(r)$ (dotted line).

FIG. 4. MD results of the circumsphere (void) radii distribution, $D(R_V)$, using the RIM (dotted line) and the PIM (solid line).

V. Bitrián et al. **FIG. 1**

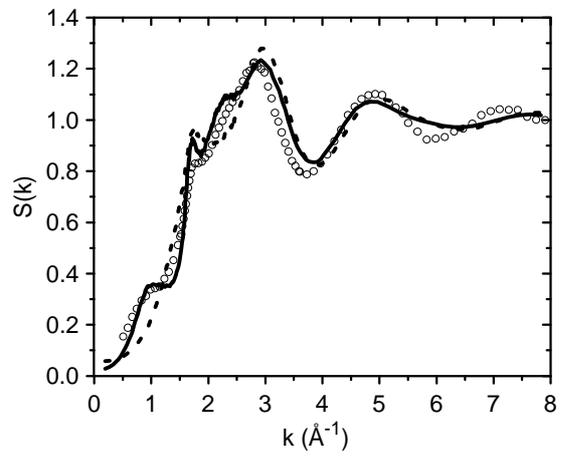

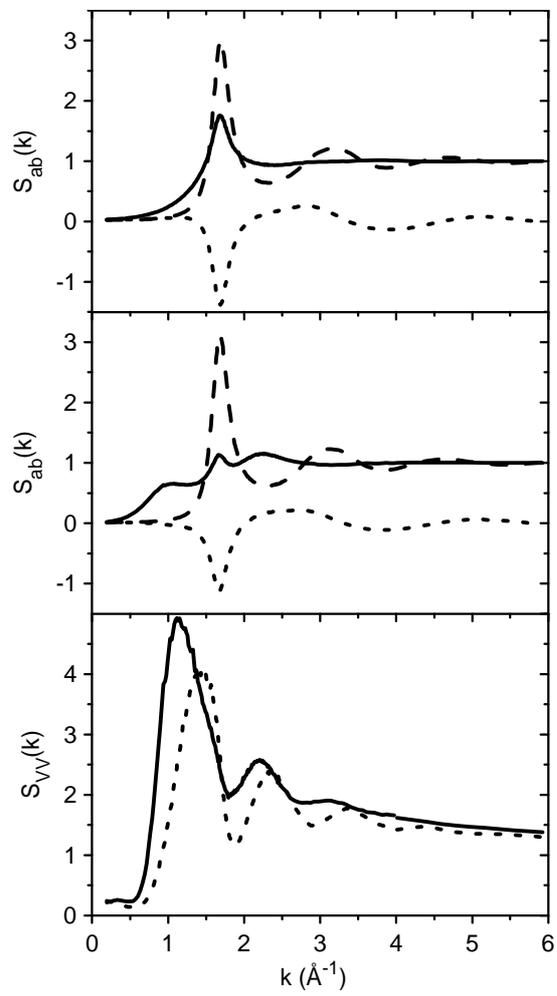

V. Bitrián et al.  **FIG. 3**

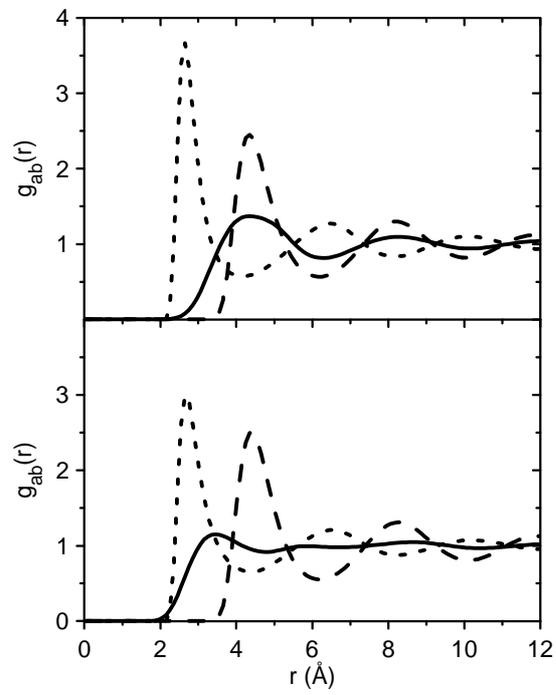

V. Bitrián et al. **FIG. 4**

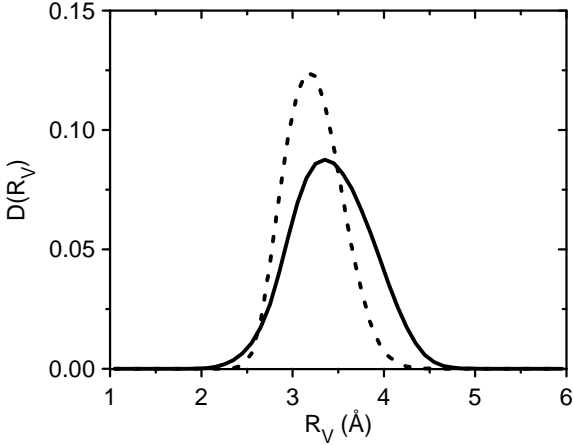